\documentclass[aps,prd,superscriptaddress,twocolumn,
floatfix,preprintnumbers,altaffilletter]{revtex4}

\usepackage{graphicx,url,amssymb,amsmath,longtable,rotating,color,units, 
wasysym,subfigure,epsfig,multirow}

\usepackage{epstopdf}
\usepackage{indentfirst}
\usepackage{latexsym}
\usepackage{amssymb}
\usepackage{mathrsfs}
\usepackage{graphicx}
\usepackage{hyperref}
\allowdisplaybreaks[1]
\begin{document}

\title{Entanglement Entropy of A Simple Non-minimal Coupling Model}


\author{Bing Sun}
\email{bingsun@mail.bnu.edu.cn}
\affiliation{Department of Physics, Beijing Normal University, Beijing 100875, China}

\author{Weizhen Jia}
\email{jiaweizhen@mail.bnu.edu.cn}
\affiliation{Department of Physics, Beijing Normal University, Beijing 100875, China}

\author{Xingyang Yu}
\affiliation{Department of Modern Physics, University of Science and Technology of China, Hefei 230026, China}

\date{\today}

\begin{abstract}
  We evaluate the entanglement entropy of a non-minimal coupling Einstein-scalar theory with two approaches in classical Euclidean gravity. By analysing the equation of motion, we find that the entangled surface is restricted to be a minimal surface. The entanglement entropy formula is derived directly from the approach of regularized conical singularity. On the other hand, by expressing Ricci scalar of the conical spacetime, we obtain the same result. In addition, we generalize the reduced geometric approach to Riemann tensor and its derivations.

\end{abstract}

\maketitle

\section{Introduction}\label{intro}
   Among the fantastic properties of black holes, gravitational entropy is one of the most impressive. After the ground-breaking work of Hawking\cite{Hawking:1974sw} demonstrated the existence of thermal radiation and temperature of black holes, it is naturally to introduce the so-called Bekenstein-Hawking entropy, which makes us treat black holes as actual thermodynamic systems. To understand the entropy of a black hole, a host of efforts have been done\cite{Bombelli:1986rw}\cite{'tHooft:1984re}  for decades, and an important step among which is using the entanglement entropy to re-understand the black hole entropy\cite{Srednicki:1993im}. Many methods have been developed after the emergence of this idea\cite{Frolov:1993ym}, including the conical singularity method\cite{Susskind:1993ws} which is adopted in this paper.  
  \par 
   
   The boom of studying holographic entanglement entropy (HEE) has started since Ryu and Takayanagi\cite{Ryu:2006bv,Ryu:2006ef} proposed the formula of HEE as
  \begin{align*}
  S_{\text{HEE}}=\frac{A_{\text{min}}}{4G_N}
  \end{align*}     
   in the context of AdS/CFT duality, where $A_{\text{min}}$ is the minimal entangled surface extending from AdS boundary to bulk. A rough proof\cite{Fursaev:2006ih} in asymptotic AdS spacetime of Einstein's theory was given within a short time. A. Lewkowycz and J. Maldacena\cite{Lewkowycz:2013nqa} verified the entanglement entropy formula in minimal coupling Einstein-scalar theory, generalized the replica trick from quantum field theory to the gravitational theory and also discussed the physical scenario behind the Ryu-Takayanagi formula. 
   
   \par 
   But if one extends the minimal coupling scalar theory to non-minimal coupling with gravity, the calculation will become much more difficult. On one hand, the scalar behaviour near the entangled surface is hard to acquire. On the other hand, integrating it with geometric quantities over the effective region is also very complicated. Besides, a general coupling could be those simple ones, such as $R\phi^2,\ R_{\mu\nu}\partial^\mu\phi\partial^\nu\phi$, or some more complicated ones, such as $R_{\mu\nu}R^{\mu\nu}(\partial\phi)^2,\ R_{\mu\nu\rho\sigma}R^{\mu\rho}\partial^\nu\phi\partial^\sigma\phi$. However, we can choose a typical example to illustrate the crucial point. Thus adding the term $\xi R\phi^2$ to the action should be a good point.

   \par 
   We mainly show how this non-minimal coupling contributes to entanglement entropy. Two equivalent approaches --- conical singularity regularization approach and reduced geometric quantities approach ---  are used here, both of which come from the conical singularity method. Dealing with equations of motion, we obtain the restriction of entangled surface, which is a minimal surface. Our analysis in this non-minimal coupling form could be naturally \textit{generalized} to the cases of linear combinations of Riemann tensor and its derivations with second approach.

\section{Method}\label{Method}
  There is more than one method to evaluate $U(1)-$geometric entanglement entropy of horizon under the Euclidean formalism. We will employ the conical singularity method \cite{Susskind:1993ws,Lewkowycz:2013nqa,Fursaev:1995ef,Carlip:1993sa,Nelson:1994na} since it can deal with the action and, a step further, the entanglement entropy directly with completely known fields and geometry. Moreover, once recognizing the AdS/CFT duality with a true asymptotic boundary, one could also calculate the holographic entanglement entropy even though the mechanism of duality generally remains unknown. Here we mainly concentrate on the black hole entanglement entropy.
\par 
  
  In \cite{Lewkowycz:2013nqa}, entanglement entropy formula has been put forward with replica trick following the same procedure in field theory\cite{Calabrese:2004eu}. Consider, in Euclidean frame, a codimension-two closed surface, denoting $\mathcal{B}$, which contains all or part of the boundary in spacetime $\mathcal{M}$ with $U(1)$ symmetry, which is also a solution of action $I$. Replicate this spacetime $\mathcal{M}$ into $n$ copies and glue the surface $\mathcal{B}$ of each copy of $\mathcal{M}$ together to form a new spacetime $\mathcal{M}_n$, which is still the solution of action $I_n$, with codimension-two surface $\mathcal{B}_n$ consisting of these copies of surfaces $\mathcal{B}$. Notice that $\mathcal{M}_n$ exists a conical singularity while $\mathcal{M}$ is smooth. Then the entanglement entropy formula is
\begin{align}\label{0}
S_{\text{EE}}&=-n\partial_n\left[\log Z(n)-n\log Z(1)   \right]|_{n=1}\,.
\end{align}
For simplicity and convenience, one could modify the formula above for a bit by adding and subtracting an off-shell functional $\log Z^{\text{off}}(n)$ whose corresponding geometry $\widetilde{\mathcal{M}}_n$ is smooth everywhere. Since the functional is off-shell, one can always set it, to $(n-1)$ order, as 
\begin{align*}
\delta \log Z(n)=\log Z(n)-\log Z^{\text{off}}(n)\approx 0\,,
\end{align*}
which implies that the linearized equations of motion hold, so that the entropy formula now should be written as \cite{Lewkowycz:2013nqa}
\begin{equation}\label{1}
S_{\text{EE}}=-n\partial_n\left[\log Z^{\text{off}}(n)-n\log Z(1)   \right]\big |_{n=1}\,.
\end{equation}
Eq.\eqref{0} and \eqref{1} are our starting point to give entanglement entropy formula and corresponding to two approaches.

\par 

  Expanding the metric near the codimension-two surface $\mathcal{B}$ with Riemann normal coordinate, the line element of $\mathcal{M}$ in the static case could be read as \cite{Fursaev:2013fta,Lewkowycz:2013nqa}
\begin{align}\label{8}
ds^2=dx^adx^a+(h_{ij}+2x^ak_{aij}+x^ax^bQ_{abij})dy^idy^j\,,
\end{align}
\par 
\noindent
where index $a$ runs from $1$ to $2$ with coordinate $x^a$ describing a 2-dimensional surface and $i,j$ runs from $1$ to $D-2$ with coordinate $y^i$ describing the surface $\mathcal{B}$. Notice that the period of $\mathcal{M}_n$ is $2\pi n$, which, as mentioned previously, means that the two-dimensional surface has conical singularity at the origin while the off-shell metric dose not. As for $\widetilde{\mathcal{M}}_n$, one only need to multiply a function before the 2-dimensional surface to roll-off the conical singularity 
\begin{equation}\label{2}
ds^2=U(r)dr^2+r^2d\tau^2+(h_{ij}+2x^ak_{aij}+x^ax^bQ_{abij})dy^idy^j\,,
\end{equation}
where we demand that $U(r)|_{r=0}=n^2$, $U(r)|_{r\rightarrow \infty}=1$ and $U(r)=1+(n^2-1)f(r,a)$ where $f(r,a)$ is any smooth function satisfying the conditions above and $a$ is a small regulator. Therefore, looking at the metric \eqref{2} and omitting the part $\mathcal{B}_n$, in the vicinity of $r=0$, we have induced line element $d\tilde{s}^2|_{r\rightarrow 0}=n^2dr^2+r^2d\tau^2$ and thus the geometry has period $2\pi n$ with $\mathbb{Z}_n$ symmetry, which smoothes the conical singularity. We could take $U(r)=1+(n^2-1)\exp\left(-r^2/a^2 \right)$ which could be separated into zeroth order and $(n-1)$-order, for example.

   \par 
   On the other hand, dealing with the Riemann tensor of regularized metric $\widetilde{\mathcal{M}}_n$ through a direct computation and expanding near $r=0$, one can get 
   \begin{align}
   R=&R_C+R_B+3k_{aij}k^{aij}+\Gamma^c_{\ ab}g^{ab}k_c+\partial_ag^{ac}k_c\nonumber\\
   &-\Gamma^a_{\ ac}g^{cd}k_d-k_ak^a+\cdots\,,
   \end{align}
where we have omitted the higher order terms because they are irrelevant to the discussion below. The label ``$C$'' denotes the geometric quantities on two dimensional regularized conical space with
\begin{align}
   R_C=\frac{U'}{rU^2}\,,
\end{align}
and ``$B$'' represents the codimension-two surface. 
These are prepared for \eqref{1}.
   
   \par 
   On the other hand, to derive the geometric quantity $R$ on manifold $\mathcal{M}_n$ rather $\widetilde{M}_n$, we could employ the fact $\lim_{a\rightarrow 0}\widetilde{\mathcal{M}}_n\rightarrow \mathcal{M}_n$. Then, the integrating of $\sqrt{g} R$ over the region $\widetilde{\mathcal{M}}_n$ can be separated into two parts --- the contribution near the tip and far away from the tip 
   \begin{align}\label{26}
   &\quad \lim_{a\rightarrow 0}\int_{\widetilde{\mathcal{M}}_n} d^Dx\sqrt{g}R\nonumber\\
   &=4\pi (1-n)\int_{\mathcal{B}} d^{D-2}\sqrt{h}+\int _{\mathcal{M}_n/\mathcal{B}_n} d^Dx\sqrt{g}R\nonumber\\
   &=4\pi (1-n)\int_{\mathcal{B}} d^{D-2}\sqrt{h}+n\int _{\mathcal{M}} d^Dx\sqrt{g}R\ \ (\text{$U(1)$\ symmetry})
   \end{align}
which can be re-expressed as 
   \begin{equation}\label{27}
   ^{(n)}R=nR+2(1-n)\delta (r)\,.
   \end{equation}
Obviously, the quantities in right hand side are defined on manifold $\mathcal{M}$ now with the conical singularity representing as a delta function which is defined as 
    \begin{equation}
    \int_{C} dr\ r\delta(r)=1\,.
    \end{equation}
Under these condition together with metric \eqref{8}, we can give a calculation \eqref{0} to obtain entanglement entropy.

\section{Model Setup}\label{Setup}

  Generally speaking, a non-minimal coupling action could be rather complicated. Nevertheless, we can pick out a simple but representative one to accomplish the calculation of entanglement entropy, and other cases could be treated analogously. Hence, we choose a relatively simple Euclidean action as follow
\begin{align}
\label{3}
I=\int d^Dx\sqrt{g}&\left[\frac{1}{2\kappa^2}(R-2\Lambda)+\frac{1}{2}\xi \phi^2 R\right.\nonumber\\
&\ \left.-\frac{1}{2}g^{\mu\nu}\partial_\mu \phi\partial_\nu\phi-V(\phi) \right]
\end{align}
with the equations of motion
\begin{align}
&\ \ (1+\xi\kappa^2\phi^2)R_{\mu\nu}-\frac{1}{2}\left[1+\kappa^2\xi(1+4\xi)\phi^2\right]g_{\mu\nu}R+\Lambda g_{\mu\nu}\nonumber\\
=&\kappa^2\left[(1+2\xi)\partial_a\phi\partial_b\phi-\left(\frac{1}{2}+2\xi\right)g_{\mu\nu}(\partial\phi)^2+2\xi\phi\nabla_\mu\nabla_\nu\phi\right.\nonumber\\
&\ \ \ \ \ \left. -g_{\mu\nu}\left(V+2\xi\frac{\partial V}{\partial\phi}\right)\right]\,, \label{4}\\
&\Box\phi+\xi R\phi=\frac{\partial V}{\partial\phi}\,. \label{5}
\end{align}
As mentioned above, geometries $\mathcal{M}_n$ and $\mathcal{M}$ satisfy these equations, but $\widetilde{\mathcal{M}}_n$ does not. Since $\delta\log Z(n)\approx -\delta I_n= 0$, however, the $(n-1)$-order linearized equations of motion are satisfied by $\widetilde{\mathcal{M}}_n$.

\section{Equation of motion analysis}\label{EOM}
\subsection{Gravitational Field Sector}
  In this subsection, we change our convention into \cite{Lewkowycz:2013nqa} and follow their discussion. The original smooth spacetime is $\mathcal{M}_n$ and $\mathcal{M}\cong\mathcal{M}_n/\mathbb{Z}_n$ has a conical singularity.
  \par 
  Before analysing equations of motion, we may change the form of the off-shell metric \eqref{2} in complex coordinate separately as \cite{Dong:2013qoa,Lewkowycz:2013nqa}
\begin{align}
  ds^2&=e^{-2A(z,\bar{z})}dzd\bar{z}\nonumber\\
  &\ \ \ +(h_{ij}+2x^ak_{aij}+x^ax^bQ_{abij})dy^idy^j\label{6}\,,
\end{align}
where 
\begin{align*}
A(z,\bar{z})&=\frac{\varepsilon}{2}\log (z\bar{z}+a^2)\,, \ \ \varepsilon=n-1
\end{align*}
and $a$, as a regulator, is a very small positive number.
\par 

  To do the perturbation, we add the $\delta g_{\mu\nu}$, $\delta \phi$ to the metric \eqref{6} and scalar equation \eqref{4}. In this way, the metric and scalar equation will be modified to linear order in $\varepsilon$ or $(n-1)$-order.  Notice that $A$ and $\delta A$ are $(n-1)$-order quantities, which should be treated as perturbation of $z\bar{z}$- and $\bar{z}z$-component, so that the perturbation $\delta g_{z\bar{z}}$ could be set as zero on the two-dimensional surface. Besides, as a gauge choice, we set $\delta g_{zz}=\delta g_{\bar{z}\bar{z}}=0$. Perturbation $\delta g_{\mu\nu}$ has apparently a periodicity of $2\pi$, i.e. $\delta g_{\mu\nu}(\tau)=\delta g_{\mu\nu}(\tau+2\pi)$.
\par 

  Now consider the linearized equation of gravitation.  We could now only focus on the divergent term in $\delta[(1+\xi\kappa^2\phi^2) R_{\mu\nu}]$ as the Ricci scalar $R$ and its variation $\delta R$ are not divergent since they are related to the quantities $T$, $\delta T$ and $\phi$, $\delta\phi$, all of which are convergent when contracting the equations of motion with $g^{\mu\nu}$. By calculating the variation of $R_{ab}$, we find that the divergent  terms potentially come from second derivative of $\delta g_{ij}$ respect to label $a$ and the derivative of function $A$ or $\delta A$ 
\begin{equation*}
(1+\xi\kappa^2\phi^2)\delta R_{ab}+\xi\kappa^2\delta(\phi^2) R_{ab}+\text{regular\ term\ as\ }r\rightarrow 0\,,
\end{equation*}
but the second term is second order which means it could be neglected. A simple calculation gives 
   \begin{align*}
   (1+\xi\kappa^2\phi^2)\delta R_{ab}&=(1+\xi\kappa^2\phi^2)(k_c\Gamma^c_{\ ab}-\frac{1}{2}\partial_a\partial_b\delta g)\nonumber\\
   &\quad +\text{regular\ term\ as\ }r\rightarrow 0\,,
   \end{align*}
where $\delta g=g^{ij}\delta g_{ij}$ and $k_a=\text{Tr}\ k_{aij}$. Thus considering the $zz$- and $\bar{z}\bar{z}$-component
\begin{align*}
\delta R_{zz}&\sim -\frac{1}{2}\partial_z^2\delta g+k_z\delta\Gamma^z_{\ zz}=-\frac{1}{2}\partial_z^2\delta g-\frac{\varepsilon}{z}k_z\,,\\
\delta R_{\bar{z}\bar{z}}&\sim -\frac{1}{2}\partial_{\bar{z}}^2\delta g+k_{\bar{z}}\delta\Gamma^{\bar{z}}_{\ \bar{z}\bar{z}}=-\frac{1}{2}\partial_{\bar{z}}^2\delta g-\frac{\varepsilon}{z}k_{\bar{z}}\,,
\end{align*}
one could derive 
\begin{equation}\label{17}
k_z=k_{\bar{z}}=0\,,
\end{equation}
which is exactly the condition for the minimal surface if we require the $U(1)-$ symmetry of $\tau$ for perturbation and convergence of $\delta R_{ab}$.

\subsection{Scalar Field Sector} \label{SFS}

    We now continue to analyse the behaviour of scalar field near the origin. Notice that the Ricci scalar $R$ could be decomposed as 
\begin{equation}\label{24}
R=R_C+R_B+3k_{aij}k^{aij}+\cdots
\end{equation}
under minimal surface condition $k_a=0$. From equation of motion of scalar, we could separate it into two parts near the origin ----- one describes scalar field on $\mathcal{B}$ and left part describes the 2-dimensional surface:
    \begin{equation}\label{18}
    \Box_1\phi+\Box_2\phi+\xi R_C\phi+\xi R_B\phi+3\xi k_{aij}k^{aij}-m^2\phi=0\,,
    \end{equation}
where $V$ is chosen as $\frac{1}{2}m^2\phi^2$ and $\Box_1$ and $\Box_2$ represent the Laplacian on 2-dimensional surface and codimension-two surface $\mathcal{B}$ respectively. With $R_C=0$ from metric \eqref{8}, adopting the method of separation of variables, it becomes two decoupled equations of motion
\begin{align}
\Box_1\psi_\lambda(r,\tau)-(\lambda^2+m^2)\psi_\lambda(r,\tau)&=0 \label{9}\,,\\
\Box_2Y_\lambda(y)+\xi R_2Y_\lambda(y)+3\xi k^2(y)Y_\lambda(y)+\lambda^2Y_\lambda(y)&=0 \label{10}\,,
\end{align}
while $\phi=\sum_\lambda\psi_\lambda(r)Y_\lambda(y)$. Under $U(1)$ symmetry condition and static condition, $\psi_\lambda(r,\tau)=\psi_\lambda(r)$ and \eqref{9} is turned into
\begin{equation}\label{11}
\psi_\lambda''+\frac{1}{r}\psi_\lambda'-(\lambda^2+m^2)\psi_\lambda=0\,,
\end{equation}
which gives the solution of modified Bessel function $I_0(\sqrt{\lambda^2+m^2}r)$ if $(\lambda^2+m^2)> 0$, or Bessel function $J_0(\sqrt{|m^2+\lambda^2|}r)$ when $(\lambda^2+m^2)< 0$, or $\psi_\lambda=C\log r+D$ when exactly $\lambda^2=-m^2$. With a physical consideration, we impose $C=0$ so that it is a trivial case.

\section{Entanglement Entropy}\label{EE}
Back to the formula \eqref{1}. With the saddle approximation, the first term is $\log Z^{\text{off}}(n)\approx -I^{\text{off}}_n$, which is the same action to \eqref{3}. Also, from the expression \eqref{1} and geometries of $\widetilde{\mathcal{M}}_n$, $\mathcal{M}$ at the limit $a\rightarrow 0$, the entropy only originates from the neighbourhood of the tip of the cone, measured by $b$, a small quantity, whose order is less than $a$, i.e. $b/a\rightarrow \infty$. All of the fields and their derivations in this action are off-shell. The Ricci scalar $R$ in metric \eqref{2} could be decomposed in the vicinity of origin as \eqref{24}. Seeing that $U'$ in \eqref{2} is $(n^2-1)f'$, the integration $\sqrt{g}\phi^2 R$ of action $I_n^{\text{off}}$ over coordinate $r$ is $(n-1)$-order with $\phi$ no lower than zeroth order. Because $U'(r)|_{n=1}=0$, the only contribution to the entanglement entropy stems from the derivative respect to $n$ acting on $U'$ if it dose not vanish as well. Hence the evaluation of this term involving field $\phi$ and function $U(r)$ can be simplified with $n=1$ except for $U'$. Furthermore, $\phi$ in $\sqrt{g}\phi^2 R$ is precisely the solution of the equations of motion which we have discussed in Sec. \ref{SFS}.
\par 
  we could know by observing the action that there is no term gives contribution to the entropy except $R_C$ in the integration since the second term in formula \eqref{1} would cancel them. The value of $\int d^Dx\sqrt{g} R$ have been shown in equation \eqref{26}. This result does not depend on the choice of regulator $a$ and function $U(r)$.
\par 
  As seen above, only terms of two derivative of $x^a$ respect to metric can give non-zero result and others would vanishes as we requiring $a\rightarrow 0$. Therefore another non-trivial contribution is the second term in the action \eqref{3}. The results appear to rely on the choice of $U(r,a)$. However, all kinds of regularization actually should give the same answer as we will discuss later.  
  \par   
  Recalling that the integration $\int\sqrt{g}\phi^2 R$ over coordinate $\tau$ contributes $2\pi n$, with $U(r)=1+(n^2-1)\exp\left(-r^2/a^2 \right)$ setted in Sec. \ref{Method} and the fact $b/a\rightarrow \infty$, known scalar field $\phi$ composed as 
  \begin{align}\label{28}
  \phi(r,\tau,y)=&\sum_{\lambda_1}A_{\lambda_1}J_0(\sqrt{|m^2+\lambda_1^2|r})Y_{\lambda_1}(y)\nonumber\\
  &+\sum_{\lambda_2}B_{\lambda_2}I_0(\sqrt{(m^2+\lambda_1^2)r})Y_{\lambda_2}(y)\nonumber\\
  &+\sum_{\lambda_3}C_{\lambda_3}Y_{\lambda_3}(y)\,,
  \end{align}
we can acquire the following result after a tedious calculation, 
  
  \begin{align}\label{13}
  &-n\partial_n\big |_{n=1}\left[\frac{1}{2}\xi\int_{\widetilde{\mathcal{M}}_n}d^Dx\sqrt{g}\phi^2 R-\frac{n}{2}\xi\int_{\mathcal{M}}d^Dx\sqrt{g}\phi^2 R\right]\nonumber\\
    =&-n\partial_n\big |_{n=1}\left[\frac{1}{2}\xi\int_{r\sim 0}d^Dx\sqrt{g}\phi^2 R\right]\nonumber\\
  =&4\pi\xi\sum_{\lambda_1,\lambda_2}A_{\lambda_1}B_{\lambda_2}J_0\left(\frac{a^2}{2}\xi_1|\xi_2|\right)\nonumber\\
  &\quad \quad \quad \quad \times \exp\left(\frac{a^2}{4}(\xi_1^2+|\xi_2|^2)\right) F_{\lambda_1,\lambda_2}\nonumber\\
  &+4\pi\xi\sum_{\lambda_1,\lambda_2}A_{\lambda_1}C_{\lambda_2} e^{-\frac{a^2}{4}|\xi_1^2|}F_{\lambda_1,\lambda_2}\nonumber\\
  &+4\pi\xi  \sum_{\lambda_1,\lambda_2}B_{\lambda_1}C_{\lambda_2}e^{\frac{a^2}{2}\xi_2^2}F_{\lambda_1,\lambda_2}\nonumber\\
  &+2\pi\xi\sum_{\lambda_1,\lambda_2}A_{\lambda_1}A_{\lambda_2}J_0\left(\frac{a^2}{2}\sqrt{|\xi_1^2||\xi_2^2|}\right)\nonumber\\
  &\quad\quad \quad \quad \times \exp\left( -\frac{a^2}{4}(|\xi
  _1|^2+|\xi_2|^2)\right) F_{\lambda_1,\lambda_2}\nonumber\\
  &+2\pi\xi\sum_{\lambda_1,\lambda_2}B_{\lambda_1}B_{\lambda_2}I_0\left(\frac{a^2}{2}\xi_1\xi_2\right)\nonumber\\
  &\quad \quad \quad \quad \times\exp\left(\frac{a^2}{4}(\xi_1^2+\xi_2^2)\right) F_{\lambda_1,\lambda_2}\nonumber\\
  &+2\pi\xi \sum_{\lambda_1,\lambda_2}C_{\lambda_1}C_{\lambda_2}F_{\lambda_1,\lambda_2}
  \end{align}
\par 
\noindent
where $F_{1,2}$ and $\xi_i$ are defined as 
\begin{align}
F_{1,2}&\equiv \int_{\mathcal{B}} \sqrt{h}Y_{\lambda_1}(y)Y_{\lambda_2}(y)dy^1\cdots dy^{D-2}\,,\nonumber\\
\xi^2_i&=({m^2+\lambda_i^2}), \ \ \ i=1,2,\nonumber
\end{align}
$Y_\lambda(y)$ is the solution of \eqref{10} and all of $A,B,C$ are expansion coefficients occur in the expansion of field $\phi$. Apparently, equation \eqref{13}  depends on regulator $a$ as we desired. Thus we can obtain the entanglement entropy on the condition of $a\rightarrow 0$
\begin{align}\label{16}
S_{\text{EE}}=&\frac{A}{4G_N}\nonumber\\
&+4\pi\xi\sum_{\lambda_1,\lambda_2}A_{\lambda_1}C_{\lambda_2} F_{1,2}+4\pi\xi\sum_{\lambda_1,\lambda_2}B_{\lambda_1}C_{\lambda_2} F_{1,2}\nonumber\\
&+2\pi\xi\sum_{\lambda_1,\lambda_2}C_{\lambda_1}C_{\lambda_2}F_{1,2}+4\pi\xi\sum_{\lambda_1,\lambda_2}A_{\lambda_1}B_{\lambda_2} F_{1,2}\nonumber\\
&+2\pi\xi\sum_{\lambda_1,\lambda_2}A_{\lambda_1}A_{\lambda_2} F_{1,2}+2\pi\xi\sum_{\lambda_1,\lambda_2}B_{\lambda_1}B_{\lambda_2} F_{1,2}\,.
\end{align}
Now, although we have obtained the entropy formula, the procedure \eqref{13} seems to depends on the choice of regularization function $U(r,a)$, and result might be divergent if one takes an inappropriate $U(r,a)$, such as $U(r,a)=\frac{r^2+n^2a^2}{r^2+a^2}$\cite{Fursaev:2013fta,Fursaev:1995ef}, which is not convergent enough comparing with the Bessel functions.
Expending $\phi$ according to Bessel functions gives
\begin{align}\label{31}
\phi(r,y)=A(y)+r^2B(y)+\cdots 
\end{align} 
However, one could expect the scalar field $\phi$ only to be expanded into the first order, just like all the other quantities. Thus, the higher order terms including $r^2B(y)$ turn out to be irrelevant terms, which implies that the contribution to the entropy is the integration
  \begin{align*}
  &\frac{1}{2}\xi\int_{r\sim 0}d^Dx\sqrt{g}A^2(y) R\\
  =&\frac{1}{2}\xi \int _{\widetilde{\mathcal{B}}_n}\sqrt{h}A^2(y) d^{D-2}y\int _0^{2\pi n}d\tau\int_0^b \sqrt{U}r dr\sqrt{g}R_C+\cdots
  \end{align*}
which is exactly the result \eqref{16}. Therefore the dependence on regularization should be a misapprehension due to higher order terms and integration over $(0,+\infty)$ (stems from $b/a\rightarrow \infty$).

\par 

  Moreover, we can also consider another approach --- reduced geometric quantities --- to work out the entropy formula \eqref{16} without analysing the divergent behaviour of the integration. The Ricci scalar has been represented into the form \eqref{27}.  A direct integration of the non-minimal coupling term is written as
  \begin{align}
  \frac{1}{2}\xi\int _{\mathcal{M}_n}\sqrt{g}&\phi^2Rd^Dx=\frac{n}{2}\xi\int _{\mathcal{M}}\sqrt{g}\phi^2Rd^Dx\nonumber\\
  &+2\pi (1-n)\xi\int_{\mathcal{B}}\sqrt{h}\phi(0,y)d^{d-2}y\,,
  \end{align}
one can easily derive the entropy formula \eqref{16} from here together with the solution $\phi(r,\tau,y)$ \eqref{28}.
  
  \par 
   We now reduce the Riemann tensor and the Ricci tensor analogously to $R$. Expanding the Riemann tensor under regularized manifold $\widetilde{\mathcal{M}}_n$ 
   \begin{align*}
R^a_{\ bcd}&=R^a_{C\ bcd}+\cdots\,,\nonumber\\
   R_{ab}&={R}_{Cab}+k_b^{\ ij}k_{aij}-Q_{ab}+k_c\Gamma^c_{\ ab} +\cdots
\end{align*}
with other components irrelevant, the $R^a_{C\ bcd}$ and ${R}_{Cab}$ are
   \begin{align*}
&R_{Crr}=\frac{U'}{rU}\,,\ \ \ 
   R_{C\tau\tau}=\frac{rU'}{U^3}\,,\nonumber\\
   &R^r_{C\ \tau\tau r}=-\frac{rU'}{U^3}\,,\ \ \ R^\tau_{C\ r\tau r}=\frac{U'}{rU}\,.
   \end{align*}
Integrating $R^a_{ bcd}$ and ${R}_{ab}$ over $\widetilde{\mathcal{M}}_n$ and letting $a\rightarrow 0$, it gives rise to \cite{Fursaev:1995ef}
   \begin{align}
   ^{(n)}R^{\mu\nu}_{\ \ \alpha\beta}&=n R^{\mu\nu}_{\ \ \alpha\beta}+(1-n)[\sum_{a,b=1}^2(n^{a\mu} n^a_\alpha)(n^{b\nu} n^b_\beta)\nonumber\\
  &\quad-(n^{a\mu} n^a_\beta)(n^{b\nu} n^b_\alpha)]\delta (r)\,,\label{29}\\
   ^{(n)}R_{\mu\nu}&=n R_{\mu\nu}+(1-n)(\sum_{a=1}^2n^a_\mu n^a_\nu)\delta (r)\,,\label{30}
   \end{align}
where the Riemann tensor and its derivations of r.h.s of two equations above are defined in the spacetime $\mathcal{M}$ and the $n^a_\mu$s are the normal vectors orthogonal to surface $\mathcal{B}$. Now we have reduced the Riemann tensor and its derivations in the conical singularity spacetime $\mathcal{M}_n$ into the conical singularity contributions and geometric quantities in the  original spacetime $\mathcal{M}$.

  \par 
    It is obviously that the reduced geometric approach is much simpler than the previous approach. In addition, the linear combinations of the geometric quantities in a non-minimal coupling theory are proportional to $(n-1)$ and thus the derivative of $n$ respect to the action $I_n$ must be acting on the geometric quantities or vanishing. Therefore all of other quantities take $n=1$ and delta functions from linear combinations of the Riemann tensor and its derivations impose other fields valued on entangled surface $r=0$.
    
   \par 
   For instance, given the non-minimal coupling term in an action 
   \begin{align}\label{22}
   I'=\frac{1}{2}\alpha\int_{\mathcal{M}} d^dx\sqrt{g}G_{\mu\nu}\partial^\mu\phi\partial^\nu\phi\,,
   \end{align}
where $G_{\mu\nu}=R_{\mu\nu}-\frac{1}{2}g_{\mu\nu}R$ is Einstein tensor, $^{(n)}G_{\mu\nu}$ could be expressed as
   \begin{align*}
   ^{(n)}G_{\mu\nu}=G_{\mu\nu}+(1-n)(\sum_{a=1}^2 n^a_\mu n^a_\nu)\delta (r)-(1-n)g_{\mu\nu}\delta (r)\,.
   \end{align*}
If we insist on $U(1)$ symmetry and the $\phi$ remains the form $\phi(r,\tau,y)=\sum_{\lambda} A_\lambda \psi_\lambda (r)Y_\lambda(y)$, it gives rise to 
    \begin{align}\label{23}
    S'_{\text{EE}}=&\pi \alpha\sum_{\lambda_1,\lambda_2}A_{\lambda_1}A_{\lambda_2}\psi_{\lambda_1}(0)\psi_{\lambda_2}(0)\nonumber\\
    &\times\int_{\mathcal{B}} d^{d-2}y\sqrt{h}h^{ij}\partial_iY_{\lambda_1}(y)\partial_jY_{\lambda_2}(y)\,.
    \end{align}
Thus we can see that the reduced geometric approach does simplify the procedure a lot.

\section{Conclusions}\label{conclusions}

  In this paper, we have derived the entanglement entropy of a non-minimal coupling theory \eqref{3}. With the procedure of analysing linearized equations of motion, we showed that the chosen entangled surface $\mathcal{B}_n$ must be a minimal surface \eqref{17}. Also, in the vicinity of $r=0$, the scalar field in the equation of motion \eqref{18} behaves as multiplying the functions on the entangled surface by Bessel functions, modified Bessel functions or constants. After a series of analyses, we conclude that the derivative of $n$ respect to non-minimal coupling term must act on Ricci scalar $R_C$ or it will vanish when $n=1$ substitutes into the final result and all quantities except $R_C$ could take the $n=1$ solution. Then the integration of the non-minimal coupling term gives the entropy formula \eqref{13}. 
\par 

  During the evaluation of the entropy, it seems that the integration \eqref{13} would depend on the specific form of smoothed regularization function. This semblance springs from two aspects: the expansion of $\phi$ into higher orders and the step of $b/a\rightarrow \infty$. Nevertheless, the integration is convergent if one integrates after $a\rightarrow 0$, since the higher order terms are removed. Therefore the sequence of the integration and the limit $a\rightarrow 0$ is significant here. 

  \par 
   By re-evaluating the entanglement entropy we have found that the same result can also be given from the reduced geometric quantities approach, which is a much simpler approach without having to analyse the divergence and calculating the integration respect to $r$ because of the appearance of the delta function $\delta(r)$. In addition, this approach could be generalized to the linear combinations of the known reduced geometric quantities --- expressed as \eqref{27} $\sim$ \eqref{30} --- with non-minimal coupling to scalar fields and we give an example \eqref{22}.
  
  \par 
  Note that the formula \eqref{16} can calculate not only  the black hole entanglement entropy but also holographic entanglement entropy if the spacetime has a true asymptotic boundary. 

  \par 
  Furthermore, our entropy formula as entanglement entropy has a natural relation with Wald entropy as thermodynamic entropy. \cite{Fursaev:1995ef} gave a easy comment on entanglement entropy and Wald entropy, while \cite{Dong:2013qoa} used Wald entropy to holographic entanglement entropy in the cases of pure geometric quantities directly. Actually, in our frame, according to saddle approximation, eq.\eqref{0} can be re-expressed as 
    \begin{align*}
    S_{\text{EE}}&=n\partial_nI_n[R_{\mu\rho\nu\sigma}]|_{n=1}-I_1[R_{\mu\rho\nu\sigma}]\\
    &=\int_{\mathcal{M}_n} d^Dx\sqrt{g}\frac{\partial I_{n}}{\partial R^{\mu\rho\nu\sigma}}\partial_nR^{\mu\rho\nu\sigma}\bigg |_{n=1}-I_1[R_{\mu\rho\nu\sigma}]\\
    &=\int_{\mathcal{B}}d^{D-2}x\sqrt{h}\frac{\partial I_{n}}{\partial R^{\mu\rho\nu\sigma}}\bigg |_{n=1}\nonumber\\
  &\quad \times \left[\sum_{a,b=1}^2(n^{a\mu} n^{a\nu})(n^{b\rho} n^{b\sigma})-(n^{a\mu} n^{a\sigma})(n^{b\nu} n^{b\rho})\right]\,.
    \end{align*}
If one identifying the finial line with $\varepsilon^{\mu\nu}\varepsilon^{\rho\sigma}$, this entanglement entropy is precisely Wald entropy and it indeed holds with a mathematical treatment under the metric \eqref{8}. Then, when considering scalar fields coupling to linear combinations of Riemann tensor and its derivations, this expression reads as
    \begin{align*}
    S_{\text{EE}}&=\int_{\mathcal{B}}d^{D-2}x\sqrt{h}\frac{\partial I_{n}[R_{\mu\nu\rho\sigma},\phi]}{\partial R^{\mu\rho\nu\sigma}}\bigg |_{n=1}\nonumber\\
  &\quad \times \left[\sum_{a,b=1}^2(n^{a\mu} n^{a\nu})(n^{b\rho} n^{b\sigma})-(n^{a\mu} n^{a\sigma})(n^{b\nu} n^{b\rho})\right]\\
  &\quad +\int _{\mathcal{B}}d^{D-2}x\sqrt{h}\  \frac{\partial I_n[R_{\mu\nu\rho\sigma},\phi]}{\partial \phi}\partial_n\phi\bigg |_{n=1}\,.
    \end{align*}
Expanding $\phi$ near the horizon with the form \eqref{31} implies zero contribution and thus it is the same as before. Consequently, the argument remains to hold in the cases of the linear combinations of the geometric quantities.

\begin{acknowledgments}
  We would like to thank Prof. Hong L\"u, Prof. Bin Zhou and Dr. Hongbao Zhang for helpful discussions.

\end{acknowledgments}

\bibliographystyle{unsrt}
\bibliography{Nonminimal-coupling}

\end{document}